\documentclass[10pt]{article}
\usepackage{url}
\usepackage[hang,flushmargin]{footmisc}
\usepackage{perpage}
\MakePerPage{footnote}

\begin{document}

\title{The myth of cell phone radiation}
\author{Vasant Natarajan\thanks{email: vasant@physics.iisc.ernet.in} \\
{\small \em Department of Physics, Indian Institute of Science, Bangalore 560 012, India}}

\date{\today}

\maketitle

\begin{abstract}
We discuss the purported link between cell-phone radiation and cancer. We show that it is inconsistent with the photoelectric effect, and that epidemiological studies of any link have no scientific basis.

\vspace{2mm}

\noindent
{\bf Keywords: } Cell phone radiation, Photoelectric effect, Cancer. \\

\end{abstract}

\noindent
{\sc Albert Einstein}, probably the greatest scientist in the history of mankind and definitely my personal hero, won the Nobel Prize not for his work on Relativity but for his explanation of the {\em Photoelectric Effect}. This explanation was considered revolutionary enough for a Nobel because it was the first independent confirmation of the paradigm-shifting quantum concept introduced by Planck a few years earlier, which stated that a light wave could only carry energy in {\em discrete} packets.

The photoelectric effect is the phenomenon where light incident on a metal (or some other surface) causes electrons to be emitted. It had been studied for quite some time before Einstein came along, and experiments had shown that photoelectrons were emitted only if the incident light had a frequency above a threshold level {\em independent of the intensity}. But the {\em number of photoelectrons} produced above threshold was indeed proportional to intensity. These observations were inexplicable from the classical wave picture of light. Assuming that there was a threshold energy that had to be overcome before electrons were emitted, one could always reach this requirement for a classical wave by suitably cranking up the intensity. But the experiments showed otherwise.

Enter Einstein and the photon picture. With the ideas that the energy per photon is quantized in units of its frequency, and that one needs a single photon with sufficient energy to produce a photoelectron, it is simple to see that there would be a threshold frequency for the effect. In addition, the number of photoelectrons would be proportional to the number of photons in the EM field, or its total energy. Thus, Einstein could explain all the observations of the photoelectric effect with the {\em reasonable} assumption that the transition involved in emitting an electron is mediated by one photon of suitable energy. It is reasonable because the transition is from one energy level where the electron is bound to another energy level where the electron is free. There are no other levels in between, which if present could be used as ``stepping stones''. This explanation is so elegant and simple to understand that it is presented to high-school students in textbooks today. But let us not forget how radical it was when it was first proposed 100 years ago. And how much of a departure from the accepted notions about light. No wonder, only a genius like Einstein could make this leap.

This is now our accepted understanding of all {\em bond-breaking} processes. Every such process involves a transition with a single photon of sufficient frequency (or energy), and a million photons of {\em sub-threshold} frequency cannot cause the transition. Or a billion. Think of it like this. If you had a cannon that could shoot a cannonball to a distance of 1 km, 10 cannons will not allow you to hit a target that is 10 km away. Cannon ranges do not add. Similarly, if you could leap a distance of 10 ft, you could jump across a stream that was 10-ft wide. But 9 additional people with the same ability cannot help you cross a 100-ft wide stream.

Which brings to the question of cell-phone radiation and its purported link to cancer. Cancer, {\em of the kind mediated by radiation}, is known to be caused by mutations in the cell-division machinery---a clear bond-breaking process---which results in uncontrolled multiplication of the cells. X-rays are well known to cause such mutations, which is why X-ray technicians are required to wear lead aprons. UV rays from the sun, those which are not stopped by the ozone layer, can cause skin cancers in people who do not have enough pigmentation to block them. That is why fair-skinned people have to use UV-blocking creams before going out into the sun. But visible light {\em cannot} cause such mutations. It is {\em sub-threshold}. And so is any EM wave whose frequency is smaller---such as infrared, microwave, radio waves, and the typical waves ($\sim 900$ MHz) used for cell phones. This means that the cell-phone photons do not have enough energy to cause a mutation in your DNA. Period. No matter what their power is---increasing their power will increase the number of photons, but they will all be below the threshold for causing cancer. They do not have enough energy to break a bond and cause a mutation. If you live next to a cell-phone transmission tower, the power levels will be higher than if you just used a cell phone, but you can be sure that all the photons are {\em harmless}.

A skeptic might argue that bond-breaking mutations are not the only way to cause cancer. True. {\em Heat} can cause damage to living tissues. And definitely if you give enough photons of sub-threshold frequency, you can heat a substance. That is why you feel hot when you go out into the sun. The sub-threshold visible and infrared photons heat up your body, \emph{but they do not cause any damage that can lead to cancer}. And the heating happens because the power density from the sun received on the surface of the earth (called the ``insolation'') is typically 1000 W/m$^2$, while that at the base of a cell-phone tower is ten thousand times smaller at about 0.1~W/m$^2$. No wonder you do not feel hot when you stand next to a cell-phone tower.

And this is exactly how a microwave oven works. It heats up the food inside by bombarding it with microwave photons. These photons have a typical frequency of 2.45 GHz, or 2.5 times that of cell phones. But even with the higher single-photon energy, the power level inside the oven required for it to heat the food is about 700 W. To understand this scale, consider that the energy inside an oven in one second is equal to that got by using a cell phone \emph{continuously for several days}. Furthermore, since a small fraction of the microwave photons come out of the oven, you actually get a larger exposure by standing next to an oven than from a cell phone. But nobody worries about it because the photons are known to be harmless, or at least nobody scared you into thinking they are harmful. Otherwise, microwave ovens would not be so commonly used today. Another figure that is useful to keep in mind is that the cell phone runs off a small battery for several days, whereas the microwave oven is the biggest electricity guzzler in the house using several kilowatts of power. The small cell phone just does not have enough energy to cause significant heating, let alone any tissue damage.

Because we evolved to live in the sunny plains of Africa, our bodies have another defense against non-ionizing radiation, namely a layer of dead cells on the outermost part of our skin. Most radiation does not make it past this layer, which is where it is absorbed to make us feel hot. Therefore, you can be sure than any radiation from the cell phone will not penetrate into the body. In addition, our brains are designed so that they do not overheat, by circulating blood as a coolant. If the bright sun cannot overheat your brain, do you think a small cell phone pressed against your ear can?

Despite the overwhelming scientific evidence that cell-phone radiation is harmless, organizations like the WHO want to play it safe and want to base their recommendations on ``epidemiological studies''---studies that compare the prevalence of cancer or other health indicators between cell phone users and nonusers. This is because there are scare-mongers who play on the fears of gullible poorly-informed people and claim that there is scientifically documented proof of such harmful effects. There was a similar unscientific claim of the hazards posed by electrical power transmission lines in the 80's and 90's. Power lines operate at a very low frequency of 50 Hz (a million times smaller than cell-phone frequencies), but have much higher power densities. The hue and cry died down only after every single epidemiological study found no link between power lines and overall health, let alone cancer. Not unexpected, {\em because there is no scientific basis for such a link to exist}. But scientists and doctors have to waste their precious time on such studies because the lay person will be satisfied only after these studies are completed.

Similar mischief-mongers told us that the radiation from computer monitors was a health risk, and then made a killing by selling ``radiation filters'' to block these rays. But most of us sit in front of a computer all day, and suffer no ill effects at all---apart from the occasional sore back that comes from bad posture and not radiation!

Such fear-mongers make truck loads of money by selling filters that block cell-phone radiation. They will give {\em anecdotal evidence} that someone who developed brain cancer ``was always talking on the cell phone'', and therefore the radiation from the cell phone {\em caused} the cancer. This is a well-known logical fallacy called {\em post hoc ergo procto hoc}, meaning that if A follows B then A was caused by B. To establish causation, the very least one must show is that no B also implies no A. And this is exactly what epidemiological studies do, they see if there is a causal difference in the prevalence of some health indicator between users and non-users. And the difference should be statistically significant, achieved by studying a large number of people and not just one or two. None has been found so far. And none will be, believe me.

In any case, all of us (cell-phone users) are unwittingly part of the largest epidemiological study ever undertaken in the history of mankind. The total number of cell-phone users in the world is now an unprecedented 80\% of the population, up by a factor of 1000 from 20 years ago. Everyone from a poor farmer in a village in India to a rich businessman in Europe uses one. Indeed, in most developed countries, the number of cell phones exceeds the population, meaning that most adults have more than one phone! {\em But there is no correspondingly large increase in the prevalence of those kinds of cancers which could be caused by cell phones (like brain tumors) during that time}. Don't you think that any ill effects of cell phones would have shown up by now in the billions of users worldwide?

We should indeed worry that our modern industrialized world is full of carcinogens---from pesticides in the food we eat, to industrial pollutants in our air and water. {\bf But cell-phone radiation is not one of them.}

\end{document}